\documentclass[twocolumn,usenatbib]{mn2e}
\usepackage{graphicx}
\usepackage{natbib}
\begin{document}
 
\title{Nonlinear $r$-modes in neutron stars: A hydrodynamical 
limitation on $r$-mode amplitudes}

\author[L.-M. Lin and W.-M. Suen]
{Lap-Ming Lin$^{1,2}$ and Wai-Mo Suen$^{2,3}$ \\
$^1$ Laboratoire de l'Univers et de ses Th\'{e}ories, 
Observatoire de Paris, F-92195 Meudon Cedex, France \\
$^2$ McDonnell Center for the Space Sciences, 
Department of Physics, Washington University, St. Louis, 
Missouri 63130, USA \\
$^3$ Department of Physics, The Chinese University of 
Hong Kong, Hong Kong, China} 

\maketitle
 
\begin{abstract} 
Previously we found that large amplitude $r$-modes could decay 
catastrophically due to nonlinear hydrodynamic effects. 
In this paper we found the particular coupling mechanism
responsible for this catastrophic decay, and identified the fluid modes involved.  
We find that for a neutron star described by a polytropic equation of state 
with polytropic index $\Gamma=2$, the coupling strength of the particular 
three-mode interaction causing the decay is strong enough that the usual picture of the $r$-mode instability with a flow pattern dominated by that of an $r$-mode 
can only be valid for the dimensionless $r$-mode amplitude less than 
$O(10^{-2})$. 

\end{abstract}


\maketitle

\section{Introduction} 

The $r$-mode instability \citep{Andersson:1,Friedman:1} in rotating neutron 
stars has attracted much attention due to its possible role in limiting the 
spin rates of neutron stars and the possibility of producing 
gravitational waves detectable by LIGO II. 
However, the significance of this instability
depends strongly on the maximum amplitude to which the $r$-mode can grow.
A number of damping mechanisms of the $r$-mode have been examined, but no 
definite conclusion of the maximum amplitude can be made 
(see \citet{Andersson:2,Friedman:2} for reviews; see also 
\citet{Andersson:3} and references therein for a discussion of recent 
developments).

In Ref. \citep{paper1}, we reported that large amplitude 
$r$-modes are subjected to a {\it hydrodynamic} instability that worked
against the gravitational-radiation driven instability.
Our numerical simulations showed that, for an $r$-mode with a large enough 
amplitude (e.g., with a dimensionless amplitude parameter $\alpha$ of order 
unity), the hydrodynamic instability operated in a timescale much shorter 
than that of the gravitational-radiation driven instability.
This hydrodynamic instability when fully developed will cause the $r$-mode to
decay catastrophically. 
As a result, the neutron star changes rapidly from the uniformly 
rotating $r$-mode pattern to a complicated differential rotating 
configuration.  However, we have not been able to 
identify the hydrodynamic mechanism responsible for this catastrophic decay.

In this paper we establish that this catastrophic decay is due mainly 
to a particular three-mode coupling between the $r$-mode and a pair of 
fluid modes.  We identify the daughter modes and show that their frequencies 
satisfy a resonance condition: the sum of their frequencies is equal to 
the $r$-mode frequency.  
The importance of three-mode coupling in the $r$-mode evolution was first 
pointed out by \citet{Arras}.  
By comparing the fully nonlinear simulation with the second-order 
perturbation equations, we determined the responsible mode coupling strength 
$\kappa$.  
The value of $\kappa$ implies that this particular three-mode coupling alone
gives a rather stringent limit to the usual picture of a neutron star with a fluid flow pattern dominated by the $r$-mode: Such a picture can only be valid with the dimensionless $r$-mode amplitude $\alpha$ smaller than or of the order $10^{-2}$. 
Beyond that, the $r$-mode must be considered to be strongly coupled to
other fluid modes.  
Instead of the gravitational radiation reaction, it is the fluid coupling 
that dominates the dynamics of the $r$-mode when its amplitude grows to 
$10^{-2}$.  Its evolution is strongly affected by the fluid coupling in a 
timescale much shorter than its growth time due to the 
gravitational-radiation driven instability.  
In particular, at that point the steady rise in amplitude of the $r$-mode due 
to gravitational radiation reaction will turn into a catastrophic decay.  
This is the main conclusion of the paper.

What happens after the catastrophic decay?  As the post-decay
fluid flow involves couplings between a large number of short wavelength modes beyond the resolution of our simulation, we cannot rule out the possibility that the $r$-mode can regain 
energy from these fluid modes and be revived within a hydrodynamic
timescale (for a detailed study of such ``bouncing'' phenomena 
see \citet{Brink05}) before the star returns to uniform rotation.
However, we do not see any hint of such bouncing in our simulations with an
$r$-mode amplitude of order unity.  This observation together with the 
consideration that a large number of fluid modes are involved in the catastrophic decay, lead us to conjecture that there is no such ``bouncing'' 
of the $r$-mode amplitude after the catastrophic decay. 

We further note that there could be fluid modes with wavelengths shorter than what our numerical simulation can resolve but with couplings to the $r$-mode strong enough that place a stronger limit on the $r$-mode amplitude, as argued by \citet{Arras} and \citet{Brink05}.  
However, the bound we establish in this paper already has significant 
implications on the astrophysical importance of the $r$-mode instability.

The study in this paper makes no particular assumption on the neutron star 
model. The star is described simply 
as a self-gravitating, perfect-fluid body with a generic 
equation of state (a polytropic EOS); no dissipative mechanism is assumed. 
We have seen the same hydrodynamic phenomenon in simulations
with stellar models of different central densities and polytropic 
indices. This suggests that the hydrodynamic mechanism being studied is 
quite generic.

\section{Numerical Results} 

\subsection{Nonlinear mode coupling}

\begin{figure}
\centering
\includegraphics[width=8cm]{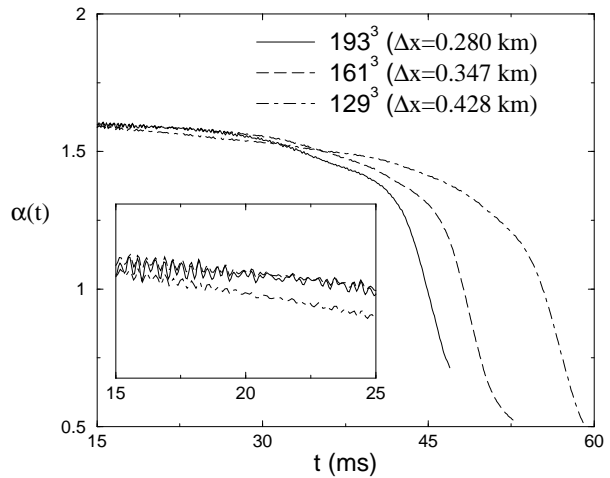}
\caption{Evolution of $\alpha$ with initial value 1.6 with three different
resolutions. The inset shows the details of the
evolution between 15 ms-25 ms.
In the $193^3$ simulation, the neutron star is represented by a total 
number of $\approx 4.4\times 10^5$ grid points.}
\label{fig:alpha1.6_new}
\end{figure}

We solve the Newtonian hydrodynamics equations for a non-viscous,
self-gravitating fluid body in the presence of a current quadrupole 
post-Newtonian radiation reaction \citep{paper1}.  
We refer the reader to \citet{paper1} for the neutron star model 
used in this study; and a detailed discussion on how we set up a large 
amplitude $r$-mode as an initial state for simulation.
To set the stage for the study in this paper, we show in 
Fig. \ref{fig:alpha1.6_new} the evolution of the $r$-mode amplitude 
$\alpha$ vs. time for the case where the initial nonlinear $r$-mode 
configuration has amplitude $\alpha=1.6$. 
We compare the results for three 
different grid resolutions $129^3$, $161^3$, and $193^3$. 
The inset in Fig. \ref{fig:alpha1.6_new} shows the details  
of the evolution in the time interval (15 ms-25 ms).  
Fig. \ref{fig:alpha1.6_new} shows that $\alpha$ starts off slowly decaying 
until $t\approx 40$ ms (in the $193^3$ simulation) at which point it decays 
catastrophically.  

\begin{figure*}
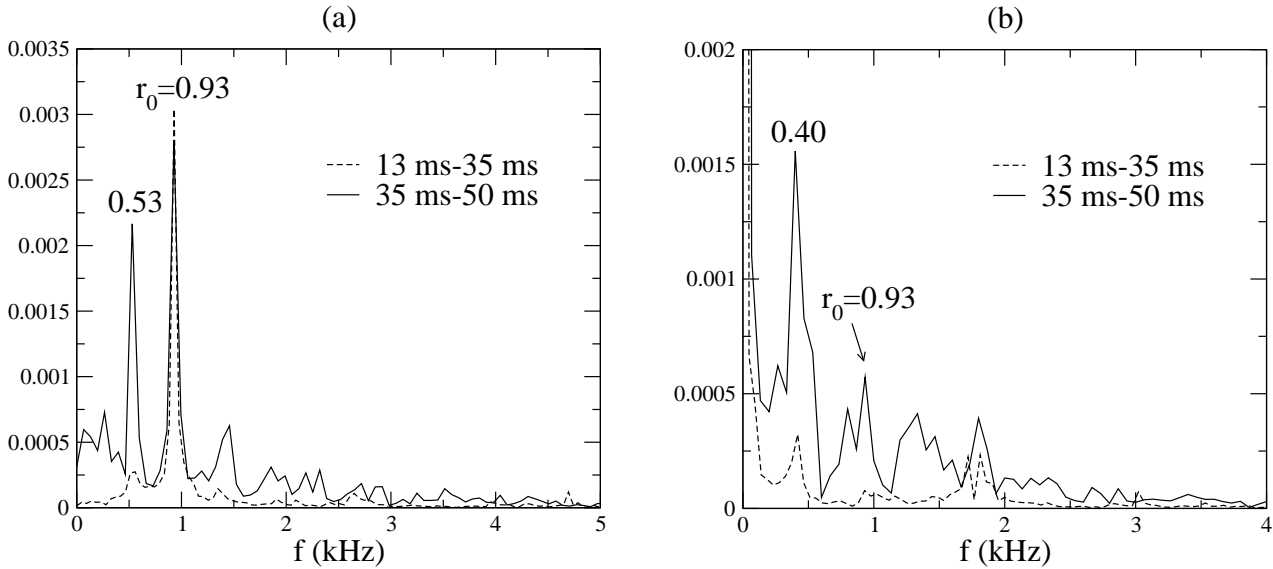

  \begin{minipage}{0.5\linewidth}
        \centering
        \includegraphics*[width=8.0cm]{fig2a.eps}
  \end{minipage}%
  \begin{minipage}{0.5\linewidth}
        \centering
        \includegraphics*[width=8.0cm]{fig2b.eps}
  \end{minipage}
    \caption{(a): Fourier spectra of $v^z$ along the $x$ axis 
		 (at $x=2.56$ km) in the evolution of Fig. 1  
		 at two time slots in the $129^3$ simulation.
		 (b): Fourier spectra of $v^y$ for the same case. }
           \label{fig:alpha1.6_129ft}
\end{figure*}

To investigate what hydrodynamic effect is responsible for the 
catastrophic decay of $\alpha$, we divide the slowly decaying phase into 
a few time intervals, and study the Fourier spectra of the velocity fields.
In Fig. \ref{fig:alpha1.6_129ft} (a), we plot the Fourier transform 
of the axial velocity $v^z$ along the $x$ axis at $x=2.56$ km inside the 
star for the $129^3$ result plotted in Fig. \ref{fig:alpha1.6_new}. 
The dashed line is the Fourier spectrum of the earlier part of the 
evolution (13 ms-35 ms), while the solid line represents that of the 
later part (35 ms-50 ms). 
We see that the dashed line has only one large peak at the $r$-mode frequency 
($r_0=0.93$ kHz).
This is compared to the spectrum at the later time slot showing many 
peaks at different frequencies. In particular, we see a significant 
peak at 0.53 kHz.

Fig. \ref{fig:alpha1.6_129ft} (b) shows the corresponding Fourier 
transforms of the rotational velocity $v^y$ along the $x$ axis at the 
same point as in Fig. \ref{fig:alpha1.6_129ft} (a). 
In contrast to Fig. \ref{fig:alpha1.6_129ft} (a), there is no 
peak at the $r$-mode frequency $r_0$ in the initial time interval 
(dashed line). 
This shows that during the initial evolution the large amplitude 
$r$-mode maintains its character as given in the linear theory:
no $v^y$ component 
on the equatorial plane. 
In the later time interval (solid line), we see various peaks, and in particular a 
significant one at 0.40 kHz.
It is also interesting to note the appearance of the peak at the 
$r$-mode frequency $r_0$ in the later time evolution. 
This is consistent with the formation of the large-scale vortex of 
the $r$-mode flow pattern during the late time evolution as reported in 
\citet{paper1}.
The $r$-mode flow pattern changes rapidly to a vortical motion on the 
equatorial plane at the time when $\alpha$ collapses rapidly. 

Figs. \ref{fig:alpha1.6_129ft} (a) and (b) demonstrate that the $r$-mode 
can leak energy to other fluid modes through nonlinear hydrodynamic couplings,
and in particular leaks energy to the fluid modes at frequencies
0.53 (henceforth the ``$A$'' mode) and 0.40 kHz (the ``$B$'' mode). 
Note that the sum of the frequencies of the $A$ and $B$ modes
exactly equals to the $r$-mode frequency ($r_0=0.93$ kHz). 
This relation suggests that the $r$-mode leaks energy to this pair of 
modes via a three-mode coupling mechanism satisfying the resonance 
condition \citep{Craik}
\begin{equation}
f_A + f_B = f_R ,
\label{eq:freq_reson}
\end{equation}
where the $f$'s are the mode frequencies.

\begin{figure}
\centering
\includegraphics*[width=8cm]{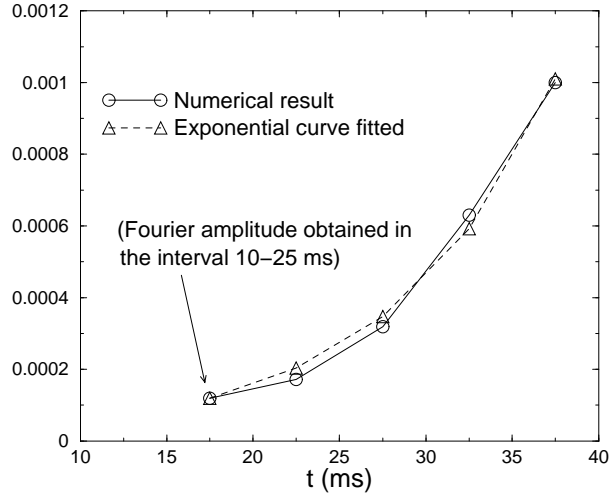}
\caption{Exponential growth of the Fourier amplitude of the $A$ mode
for the case $\alpha=1.2$ with resolution $193^3$.}
\label{fig:alpha1.2_193growth_vz}
\end{figure}

How do the coupled fluid modes grow? Instead of just two time slots, we can
further divide the slowly decaying phase of $\alpha$ into several
subintervals and perform Fourier transform on each of them, and
monitor the Fourier amplitudes of different time intervals.
In Fig. \ref{fig:alpha1.2_193growth_vz}, we plot the Fourier amplitude
of the $A$ mode vs time for the case $\alpha=1.2$ with resolution $193^3$.
In the figure, the circles connected by a solid line are numerical
data.
The first data point corresponds to the Fourier amplitude obtained in
the time interval 10 ms-25 ms. The time value of the data point (17.5 ms)
is taken to be the middle of this interval.
The second data point corresponds to 15 ms-30 ms, etc.
The dashed line is fitted to the exponential function
$f=A_0\exp( (t-17.5)/9.35 )$, where $t$ is in millisecond and
$A_0$ is the Fourier amplitude obtained in the interval 10 ms-25 ms.
Fig. \ref{fig:alpha1.2_193growth_vz} shows that the $A$ mode grows 
exponentially with a time constant $\approx 9.35$ ms during the slowly 
decaying phase of $\alpha$.
The $B$ mode grows exponentially with a time constant about 7 percents 
larger as determined numerically.  

Our results thus show that the coupled fluid modes are unstable and grow
exponentially due to their couplings to the $r$-mode.
An important point that we observed in all cases studied is that when the 
amplitudes of these two coupled daughter modes grow to a point 
comparable to that of the $r$-mode, the catastrophic decay of the 
$r$-mode sets in.

\subsection{Determination of the Decay Rate of $\alpha$}
\label{sec:decay_rate}

In the slowly decaying phase of $\alpha$, there are two mechanisms
that lead to the decay of $\alpha$ in our numerical simulation: 
(i) numerical damping arising 
from finite differencing error, which depends on the resolution
used in the simulation; and (ii) physical damping due to the nonlinear
interactions of the $r$-mode to other fluid modes, which 
depends on $\alpha$ at that moment.
Fig. \ref{fig:alpha1.6_new} shows that numerical damping is negligible 
during the early part (15 ms-30 ms) of the evolution at high enough 
resolutions. The $161^3$ and $193^3$ results agree to high accuracy. 
We can thus regard the rate of change of the $r$-mode amplitude 
$d\alpha/dt$ in the $193^3$ simulation as a reasonable approximation to its
physical value (when $\Delta x \rightarrow 0$) due to the particular
three-mode coupling seen in our simulation. 
A linear curve-fit yields 

\begin{equation}
{d\alpha\over dt} =-2.9\times 10^{-3}\ ({\rm ms})^{-1} .
\label{eq:dalpha_dt}
\end{equation}

\subsection{Determination of the waveforms} 
\label{sec:mode_char}

\begin{figure*}
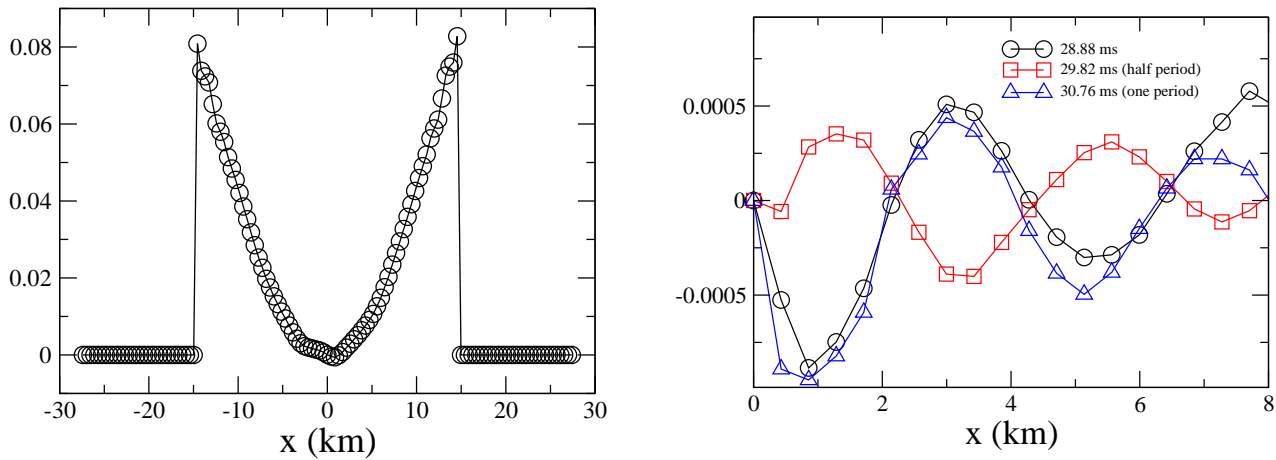

  \begin{minipage}{0.5\linewidth}
        \centering
        \includegraphics*[width=8.0cm]{fig4a.eps}
  \end{minipage}%
  \begin{minipage}{0.5\linewidth}
        \centering
        \includegraphics*[width=8.0cm]{fig4b.eps}
  \end{minipage}
    \caption{(a): The $v^z$ velocity profile along the $x$ axis at $t=28.88$ ms for
                   the $129^3$ simulation shown in Fig. \ref{fig:alpha1.6_new}. 
             (b): Waveforms of the $A$ mode for the same case. 
                   Plotted are the profiles of $\Delta v^z$ (see text) along
                  the $x$ axis at three different times. The latter two time values 
                  differ from the first one respectively by half and one oscillation 
                  period of the $A$ mode.}
           \label{fig:modeA_waveform}
\end{figure*}

To further characterize the daughter modes, we would like to determine their waveforms in addition to their mode frequencies.  However, this turns out to be non-trivial.  

In Fig. \ref{fig:modeA_waveform} (a), we show the $v^z$ velocity profile for the $129^3$ simulation shown in Fig. \ref{fig:alpha1.6_new} along the $x$ axis at $t=28.88$ ms.  We note that this is basically the velocity profile of the $m=2$ $r$-mode.  To extract the $v^z$ waveform of the $A$ mode, we need to subtract out the $r$-mode waveform. However, the $r$-mode waveform is known only in the linear approximation, while the $r$-mode being studied in this paper is not small and there can be non-linear correction to the $r$-mode waveform. To do the subtraction, we note that the non-linear correction should not change the symmetry of the $r$-mode, and in particular that its $v^z$ is symmetric with respect to reflection about the origin on the $x$ axis. We thus define the quantity 
$\Delta v^z := ( v^z(x) - v^z(-x) )/2$ on the $x$ axis. 
In Fig. \ref{fig:modeA_waveform} (b), we plot $\Delta v^z$ in the core region of the star at three different times for the $129^3$ simulation shown in 
Fig. \ref{fig:alpha1.6_new}: $t=28.88$ ms, $t=29.82$, and $t=30.76$ ms. 
The latter two time values differ from the first one respectively by 
half and one oscillation period of the $A$ mode (recall that $f_A=0.53$ kHz). 
We see explicitly that the waveform oscillates with the frequency of the $A$ mode. 
The two lines at $t=28.88$ and $30.76$ ms match pretty well, except in the region near the surface of the star where the $r$-mode velocity is large. This suggests that the extracted waveform is indeed the eigenfunction of the $A$ mode.  Note that the amplitude of the $A$ mode is less than $1\%$ of that of the $r$-mode in this early slowly-decaying phase.
The extraction would not be possible without taking advantage of the symmetry.

\begin{figure}
\centering
\includegraphics*[width=8cm]{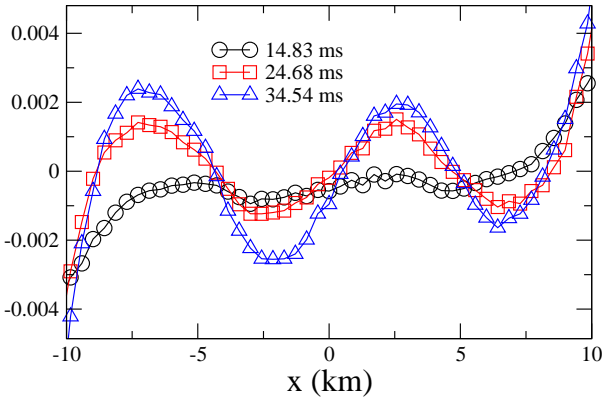}
\caption{Differential rotation pattern $\Delta v^y$ (see text) for the 
same simulation as Fig. \ref{fig:modeA_waveform}.}
\label{fig:alpha1.6_delta_vy}
\end{figure}

Next we turn to the $B$ mode, which is a mode with fluid flow in $v^y$.  One might 
expect the waveform extraction to be easier, as the $r$-mode has no $v^y$ component.  
To begin, we subtract the background rotation profile from the velocity component $v^y$ by 
defining $\Delta v^y := v^y(x) - v^y_0(x)$ along the $x$ axis, where $v^y_0$ is the 
profile at the starting point of the slowly decaying phase of $\alpha$. 
In Fig. \ref{fig:alpha1.6_delta_vy}, we plot $\Delta v^y$ along the $x$ axis at 
various times in the slowly decaying phase for the same $129^3$ simulation as before. 
We see a small differential rotation pattern growing up to an amplitude a few percent of that of the $r$-mode. Four features of this fluid pattern are: 1. This differential rotation mode has a long period, much longer than that of the $B$ mode. 2. The amplitude of this mode saturates at around $35$ ms. 3. The energy of this mode is small comparing to the $r$-mode. 4. This mode is antisymmetric with respect to the origin along the $x$ axis. 

We uncovered a differential rotation pattern unrelated to the $A$ and $B$ modes.  
What is its significance? We make two observations: 1. One might suspect that this differential rotation pattern is part of the non-linear $r$-mode profile.
It has been argued that differential rotation can be induced by second-order $r$-mode 
motion (e.g., \citet{Rez00,Sa04}) in perturbation analyses. However, the differential rotation pattern shown in Fig. \ref{fig:alpha1.6_delta_vy} does not match those
of \citep{Rez00,Sa04}. Further, we see that the differential rotation given in 
Fig. \ref{fig:alpha1.6_delta_vy} is growing in amplitude during the slowly decaying 
phase of the $r$-mode. This suggests that it may not be part of the $r$-mode. 2. 
This differential rotation pattern is different from the one that developed during 
the catastrophic decay of the $r$-mode as we reported previously \citep{paper1}.  
Further investigations of this unknown mode ($U$ mode) are needed.  

\begin{figure}
\centering
\includegraphics*[width=8cm]{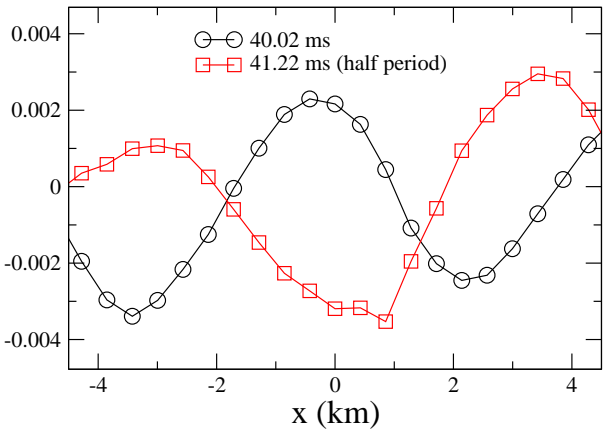}
\caption{Waveforms of the $B$ mode (see text) along the 
$x$ axis at two different times (with a time interval half the oscillation 
period of the $B$ mode) for the same simulation as Fig. \ref{fig:modeA_waveform}.}
\label{fig:modeB_waveform}
\end{figure}

To extract the $B$ mode, we subtract the antisymmetric pattern of this ``$U$ mode'' 
and find a symmetric (with respect to reflection about the origin) $v^y$ component 
with a period the same as that of the $B$ mode. It is smaller than the $U$ mode at earlier time, but overtakes the $U$ mode in amplitude at around $40$ ms. 
It keeps on growing exponentially to an amplitude comparable to that of the $r$-mode 
at the point of the catastrophic decay. 
In Fig. \ref{fig:modeB_waveform} we show the waveforms of the $B$ mode 
(obtained by subtracting the $U$ mode pattern from the profile of $\Delta v^y$) 
at around the time it becomes the dominant pattern. The two time slices $40.02$ ms 
and $41.22$ ms are separated by half of the oscillation period of the $B$ mode, and 
indeed we see that the two lines are off by $\pi$ in phase.

To summarize, we found an antisymmetric waveform in $v^z$ for the $A$ mode, and a symmetric waveform in $v^y$ for the $B$ mode. The wavelength of both modes are about 4 km in the central region of the star. Next we want to investigate if the symmetry properties of the waveforms of the modes are expected. We note that under a $\pi$-rotation 
(i.e., $\phi\rightarrow \phi + \pi$) on the equatorial plane, the Cartesian 
components of the velocity field of a toroidal fluid 
mode\footnote{Toroidal modes are characterized by fluid displacement 
vectors proportional to $\vec{r}\times \nabla Y_{lm}$, where $Y_{lm}$ 
is the standard spherical harmonics.
The $r$-mode is known to be a torodial mode. We assume that the two daughter 
modes seen in our simulations are also torodial modes (to leading order) 
because of their low frequencies compared to spheroidal modes 
(e.g., the $f$- and $p$-modes). The low frequency (spheroidal)
$g$-modes do not exist in our isentropic simulations. But note that the 
inertial modes of isentropic stars in general have a ``hybrid'' nature
(see \citet{Andersson:2} for discussion).} 
transform according to $v^x\rightarrow (-1)^{m+1} v^x$, 
$v^y\rightarrow (-1)^{m+1}v^y$, and $v^z\rightarrow (-1)^m v^z$, where 
$m$ is the azimuthal mode number (see Appendix \ref{sec:appendix}).  
This implies that the profiles of $v^x$ and $v^y$ along the $x$ axis for an $m=-1$ 
mode are symmetric with respect to reflection about the origin; while the profile of 
$v^z$ for an $m=-1$ mode is antisymmetric. We note that the $A$ and $B$ modes having 
$m=-1$ is consistent with the picture of three-mode coupling.  
The $r$-mode has $m=2$ and the conservation of angular momentum requires that the 
azimuthal mode number $m$ of the three coupled modes must 
satisfy $m_R+m_A+m_B=0$ \citep{Arras}. 

\section{Second-Order Perturbation Analysis}

\subsection{Three-mode coupling}

The numerical results indicate that the dominant hydrodynamic contribution to the decay of the 
$r$-mode is due to its coupling to the $A$ and $B$ modes. 
To gain further insight into the hydrodynamic evolution during the early part of the numerical evolution, we compare the fully nonlinear simulation to a second-order perturbation analysis.  In particular, we want to estimate the 
coupling strength of the three-mode coupling found, making use of the decay 
rate of $\alpha$ determined in Eq. (\ref{eq:dalpha_dt}).  
The second-order perturbation analysis enables us to extract information 
about the mode coupling independent of the specific value of $\alpha$ used 
in the numerical simulation.  Through this we will then be able to determine 
the behavior of the $r$-mode at $\alpha$ values smaller than that used in 
the simulations.
Note that the following analysis only makes use of information 
obtained in the slowly decaying phase of $\alpha$, where
a second-order perturbation analysis is assumed to be valid. 
In particular, it does NOT depend on the existence of a 
catastrophic decay.  The perturbation description is NOT expected to be valid when the evolution is getting near to the point of the catastrophic decay and beyond.  

In a reference frame co-rotating with a rotating star, 
the second-order perturbation equations for a system of three modes 
in resonance can be written as \citep{Schenk,Arras}
\begin{equation}
{d q_{R}\over dt} + i \omega_{R} q_{R} = {q_{R}\over \tau_{R} }
+ i \omega_{R}\kappa_{RAB}^* q_{A}^* q_{B}^* , 
\label{eq:mode_eq1}
\end{equation}
\begin{equation}
{d q_{A}\over dt} + i \omega_{A} q_{A} = {- \gamma _A q_{A} }
+ i \omega_{A}\kappa_{RAB}^* q_{R}^* q_{B}^* ,
\label{eq:mode_eq2}
\end{equation}
\begin{equation}
{d q_{B}\over dt} + i \omega_{B} q_{B} = {- \gamma _B q_{B} }
+ i \omega_{B}\kappa_{RAB}^* q_{R}^* q_{A}^* ,
\label{eq:mode_eq3}
\end{equation}
where the $q$'s represent the dimensionless complex amplitudes of the modes; 
the $\omega$'s are the rotating frame angular frequencies of the modes;
$\tau_R$ is the growth time of the $r$-mode due to 
radiation reaction; $\kappa_{RAB}$ is the coupling strength;
$Z^*$ denotes the complex conjugate of $Z$; $\gamma _A$ and $\gamma _B$ are the damping 
coefficients of the two coupled modes representing the outflow of energy from the three 
mode system. 

The system of equations (\ref{eq:mode_eq1})-(\ref{eq:mode_eq3}) gives rise to a 
resonant coupling for modes satisfying $\omega_A+\omega_B+\omega_R\approx 0$. 
Note that this is not inconsistent to the resonance relation (\ref{eq:freq_reson}) 
satisfied by the (inertial frame) frequencies of the coupled modes. 
The rotating frame angular frequency of a fluid mode $\omega$ is related to 
the inertial frame angular frequency $\omega_0$ by $\omega = \omega_0 + m\Omega$, 
where $\Omega$ is the angular velocity of the star. For the star model
($\Omega = 4.87\ ({\rm ms})^{-1}$) and $m=2$ $r$-mode with the inertial 
frame frequency $r_0=0.93$ kHz being studied, we have 
$\omega_R=3.9\ ({\rm ms})^{-1}$. Note that we define $\omega_0$ to be negative
for the $r$-mode (see \citet{paper1}). 
The rotating frame angular frequencies of the $m=-1$ daughter modes are 
respectively $\omega_A = -1.54\ ({\rm ms})^{-1} \approx -0.4 \omega_R$ and 
$\omega_B = -2.36\ ({\rm ms})^{-1} \approx -0.6 \omega_R$, suggesting that the three
modes indeed satisfy the resonance condition 
$\delta\omega=\omega_A+\omega_B+\omega_R\approx 0$.

The meaning of the $\gamma$'s used here is somewhat different from 
that of \citet{Arras} and \citet{Brink05}.  The neutron star EOS used in this paper is that 
of a perfect fluid with no viscosity.  The $\gamma$'s describe the couplings of this 
particular three-mode system to the other fluid modes of the neutron star and is a function 
of the amplitude of these other modes.  
The description of Eqs. (\ref{eq:mode_eq1})-(\ref{eq:mode_eq3}), is meaningful only when 
the $\gamma$'s are not changing in a time scale short compare to that of the dynamical 
time scales of the $A, B$ and $r$ modes, a point we will return to later in the paper.

A general expression for $\kappa_{RAB}$ is given in \citet{Schenk}.
In writing the coupled equations above, we omit the coupling
terms proportional to $\kappa_{RAA}$ and $\kappa_{RBB}$,
which are zero due to a $z$-parity selection rule \citep{Schenk}.
This selection rule states that $\kappa_{ABC}=0$ for an odd number of
$z$-parity odd modes. Since the $r$-mode has odd $z$-parity, 
$\kappa_{RAA}=0$ regardless of the $z$-parity of mode $A$.
In order that $\kappa_{RAB} \neq 0$, modes $A$ and $B$ should
have different $z$-parity.
We see in our simulations that one of the coupled modes ($A$) appears only 
in the Fourier spectrum of $v^z$, while the other one ($B$) appears only in 
the $v^y$ spectrum. 
 
Note that the definition of $\kappa_{RAB}$ used by us is the same as \citet{Arras}, but 
different from that used by \citet{Brink05} (and also \citet{Schenk}). 
In our notation, $\kappa_{RAB}$ is the ratio of the nonlinear interaction energy to the 
energy of the mode (both at unit amplitude). 
The conversion from our $\kappa$ to the one used by \citet{Brink05} is
$\kappa \rightarrow E_{\rm unit} \kappa$, where $E_{\rm unit}= {1\over 2} M R^2\Omega^2$ is 
the mode energy at unit amplitude in the normalization used by them.

In order to relate $q_R$ to the nonlinear $r$-mode amplitude defined 
in our numerical simulations we take $|q_{R}(t)| \simeq \alpha(t)$  
(the critical amplitude $\alpha_{\rm crit}$ given below applies 
as long as $|q_{R}|$ is proportional to $\alpha$).
Using Eq. (\ref{eq:mode_eq1}) and its complex conjugate, it can be shown
that $\alpha$ satisfies
\begin{equation}
{d\alpha\over dt} = {\alpha\over \tau_R} + {\omega_{R}\over \alpha}
{\rm Im}(\kappa_{RAB} q_A q_B q_R) ,
\label{eq:3mode_dalpha_eq}
\end{equation}
where ${\rm Im}(Z)$ denotes the imaginary part of $Z$. 
In the slowly decaying phase of $\alpha$,
$|q_A|\simeq |q_B|=q$ is less than $|q_R|$ (with the catastrophic 
decay sets in at $q\approx \alpha$). 
We obtain the following order-of-magnitude estimate for the 
coupling strength $|\kappa_{RAB}|$:
\begin{equation}
|\kappa_{RAB}| \sim {1 \over \omega_{R} q^2} \left| {d\alpha\over dt}
 - {\alpha\over \tau_{R}} \right|  
\sim  3\times 10^{-4} ,
\label{eq:kappa}
\end{equation}
where we have used $\alpha=1.6$, 
$d\alpha /dt = -2.9\times 10 ^{-3}$ ${\rm (ms)}^{-1}$,
$\omega_R=3.9$ ${\rm (ms)}^{-1}$, $\tau_{R}=40$ s, and 
$q \approx \alpha$. 
Note that using a smaller value of $q$ will only increase the value 
of $|\kappa_{RAB}|$. In this sense, the $|\kappa_{RAB}|$ given by 
Eq. (\ref{eq:kappa}) is a lower bound. A larger $|\kappa_{RAB}|$ gives
a tighter bound to the $r$-mode amplitude at the point of 
catastrophic decay $\alpha_{\rm crit}$ as shown below. 

To check that the value of $|\kappa_{RAB}|$ does not depend 
sensitively on the initial value of $\alpha$ used in the simulation, we have 
repeated the above calculation with a different initial $r$-mode amplitude 
$\alpha=1.2$. 
In that case, we obtain $d\alpha/dt=-1.7\times 10^{-3}\ ({\rm ms})^{-1}$ 
during the slowly decaying phase of $\alpha$. The value of $|\kappa_{RAB}|$
changes only by a few percents (within numerical error) comparing to the case 
$\alpha=1.6$.

To complete the description of the system using 
Eqs. (\ref{eq:mode_eq1})-(\ref{eq:mode_eq3}), one would also like to 
determine from the numerical simulation the values of $\gamma_A$ and 
$\gamma_B$.  
In our present setup, $\gamma_A$ and $\gamma_B$ determine the rate of energy 
leaking out of the three-mode system through coupling to other fluid modes.  
However, due to the constraint of computational resources 
we were not be able to obtain the precise values of $\gamma_A$ and $\gamma_B$ 
since they correspond to a timescale orders of magnitude longer than 
10 ms (the timescale of the growth of the daughter modes in our simulations), 
in the regime where the daughter modes $A$ and $B$ are small compared to 
the $r$-mode.

In our simulations, when the amplitudes of the two daughter modes become 
comparable to that of the $r$-mode, the perturbation equations 
(\ref{eq:mode_eq1})-(\ref{eq:mode_eq3}) are no longer valid.  
The evolution becomes fully nonlinear, and the $r$-mode decays 
catastrophically at that point.

\subsection{Amplitude of the $R$-mode at the catastrophic decay}

Now we consider how an $r$-mode would evolve from a small 
initial amplitude to the decay point in a star described by a 
perfect-fluid EOS, based on the fully nonlinear numerical simulation and 
information from the second-order perturbation analysis.
We assume that during the early evolution the $r$-mode couples mainly only 
to the two resonant modes ($A$ and $B$).
In this phase, the system can be described by Eqs.
(\ref{eq:mode_eq1})-(\ref{eq:mode_eq3}). 
The $A$ and $B$ modes are negligible initially; the $r$-mode 
grows due to radiation reaction. When $\alpha$ becomes large enough, 
$|q_A|$ and $|q_B|$ grow rapidly.  
The point at which the two daughter modes catch up in amplitude with the 
$r$-mode can be estimated by setting $d\alpha /dt=0$ in 
Eq. (\ref{eq:3mode_dalpha_eq}).  
In particular, the amplitude of the $r$-mode at that point is given by  
\begin{equation}
\alpha_{\rm crit}  \sim  {1\over \omega_R \tau_R|\kappa_{RAB}| } 
 \sim  0.02,
\label{eq:alpha_sat}
\end{equation}
where we have taken $|q_A|\approx |q_B|\approx \alpha_{\rm crit}$. 

At this point, the usual picture of a neutron star with fluid flow dominated 
by an $r$-mode pattern is no longer valid.  It is somewhat surprising that 
a single three-mode coupling of purely hydrodynamic nature could put a limit 
as strong as $\alpha \sim 0.02$.  
To the best of our knowledge this is the first explicit demonstration that 
an $r$-mode cannot be the dominant feature of a rapidly rotating neutron 
star with an amplitude beyond $\alpha \sim 0.02$.

The critical amplitude $\alpha_{\rm crit}$ 
in Eq. (\ref{eq:alpha_sat}) is defined to be the amplitude of the $r$-mode
when nonlinear hydrodynamic couplings dominate the gravitational radiation 
reaction in driving the $r$-mode evolution; it is the time when the daughter 
modes catch up in amplitudes with the $r$-mode. It should be noted that 
this critical amplitude is different from the so-called parametric 
instability threshold (see, e.g., Eq. (2) of \citet{Brink05}), which is 
defined to be the amplitude of the parent mode at the point when the two 
daughters become unstable due to nonlinear coupling and start to grow 
exponentially. 
The scalings of the two amplitudes are different. In particular, the parametric 
instability threshold does not depend on the growth time $\tau_R$ of the 
$r$-mode, while our critical amplitude $\alpha_{\rm crit}$ is inversely
proportional to $\tau_R$.

\subsection{Beyond the Critical Point}

What happens after the point $\alpha = \alpha_{\rm crit}$ is reached?  
Our numerical simulation shows that when the two daughter modes catch up 
in amplitude with the $r$-mode near $\alpha_{\rm crit}$, the following happen: 1. Many other fast growing fluid modes enter the picture.  2. The fluid flow can no longer be approximately described by the three-mode perturbation system Eqs. (\ref{eq:mode_eq1})-(\ref{eq:mode_eq3}). 3. The $r$-mode decays catastrophically.  4. After the catastrophic decay, the neutron star has a complicated differential rotating flow pattern with short length scales, and 5. there is no sign of the $r$-mode re-gaining its energy from the many fluid modes it directly or indirectly coupled to.

One important implication of this is: The maximum amplitude $\alpha_{\rm max}$ an $r$-mode can get to through the gravitational-radiation driven instability is given by $\alpha_{\rm max} = \alpha_{\rm crit}$.  However, there is a caution: our simulation is carried out with $\alpha$ of order unity but not directly at $\alpha \sim 0.02$.  There is a possibility that even though the fluid flow is no longer dominated by that of the $r$-mode after the two daughter modes catch up at $\alpha_{\rm crit} \sim 0.02$, the three-mode perturbation system Eqs. (\ref{eq:mode_eq1})-(\ref{eq:mode_eq3}) could remain a valid description of the neutron star.  The perturbation system can fail to represent the flow of the star for two reasons: (i) the mode amplitudes become large, so that higher order perturbation terms in the equations cannot be neglected, or (ii) the effects of modes other than the three modes studied become significant (this shows up also as the $\gamma$'s are changing on a short time scale); such a situation is strongly suggested by our simulation.  Unfortunately simulations of timescales long enough to accurately capture the behavior of $r$-modes with a small $\alpha$ value require a computational resource beyond what is available to us, and we cannot investigate the catastrophic decay at small $\alpha$ value in full.
 
In case the second-order perturbation equations (\ref{eq:mode_eq1})-(\ref{eq:mode_eq3}) are still valid at the point $\alpha = \alpha_{\rm crit}$, what could happen subsequent to the catastrophic decay?   
In particular, can the amplitude of the $r$-mode evolve to a value 
substantially larger than $\alpha_{\rm crit}$?  
The subsequent evolution described by Eqs. 
(\ref{eq:mode_eq1})-(\ref{eq:mode_eq3})
depends strongly on the relative values of $\gamma_A$, $\gamma_B$, $\tau_R$, 
and the frequency detuning $\delta\omega=\omega_A+\omega_B+\omega_R$.
As discussed in \citet{Arras}, a stable equilibrium solution of Eqs. 
(\ref{eq:mode_eq1})-(\ref{eq:mode_eq3})
exists if both the damping and detuning are large:
$\gamma_A+\gamma_B > \tau_R^{-1}$; 
$|\delta\omega| > (\gamma_A+\gamma_B)/2$. Otherwise, the system is 
unstable.  (Note that in our study $\gamma_A$ and $\gamma_B$ represent the rate of energy leaking out of the three-mode system to other fluid modes of the neutron star, which is described by a perfect fluid EOS with no viscosity.)

In the case where a stable equilibrium solution exists, the system will 
settle into a steady state with the mode amplitudes being roughly the same. 
The maximum amplitude of the $r$-mode $\alpha_{\rm max}$ would be given by $\alpha_{\rm crit}$.  On the other hand, in the unstable case Eqs.
(\ref{eq:mode_eq1})-(\ref{eq:mode_eq3}) predict that after the 
amplitudes of the three coupled modes become comparable, the amplitudes of 
the three modes will fluctuate in a short timescale with rapid energy 
transfer between them, and the average amplitude of all three modes could 
grow until the second-order perturbation analysis becomes 
invalid.  However, such oscillation of energy between the three modes is 
possible only if their phase relations are well preserved in the evolution, 
a condition we believe may not happen in realistic nonlinear hydrodynamic 
evolutions: at $\alpha \approx \alpha_{\rm crit}$ a large number of other fluid modes enter into the picture and affect the fluid evolution significantly. 
The energy originally in the $r$-mode is transferred to many fluid modes.  The involvement of a large number of modes suggests that 
the oscillation of energy back to the $r$-mode in a short dynamical timescale 
is unlikely. 
As an example, in a chain of $N$ coupled nonlinear oscillators
an equipartition of energy among the coupled modes is achieved if the 
energy of the initial large scale mode is larger than a certain threshold, 
which tends to zero sufficiently fast at increasing $N$ \citep{Casetti}. 
This suggests that, after cascading its energy down to a sea of coupled 
modes, a generic initial large scale mode cannot regain its energy within 
dynamical timescale. 
We believe that this is also the case for nonlinear couplings of the $r$-mode 
to other fluid modes in rotating neutron stars. 
Hence, we {\it conjecture} that such oscillation of energy cannot occur beyond 
the catastrophic decay and the amplitude of the $r$-mode is bounded by 
$\alpha_{\rm crit}$.

Nevertheless, we do not rule out the possibility that the global $r$-mode could grow 
again in a secular timescale ($\sim 40$ s) due to the driving of gravitational radiation reaction. 
We have extended some of our simulations for a timescale of 10 ms after 
the catastrophic decay. However, we do not see any evidence of the regrowth of an 
unstable $r$-mode, even with an artificially enlarged radiation reaction force. 
One possibility is that the strong differential rotation developed during the 
catastrophic decay (see \citet{paper1}) may act to stabilize the $r$-mode. If this 
is the case, it will take even a much longer time, until the differential rotation is damped 
out by viscosity (not modelled in our simulations), before the $r$-mode can grow again.
To model such regrowth scenario, if it occurs at all, would be too expensive computationally 
for present-day technology. This is still an open issue and remains to be investigated in the future.

\section{Comparison with other work}

There are two other numerical simulations of 
nonlinear $r$-modes, both finding that, large amplitude $r$-modes
can exist for a long period of time. \citet{Ste01} performed relativistic 
simulations of the $r$-mode on a fixed spacetime background. 
They started with a large $r$-mode perturbation in a relativistic 
rotating star and evolved the system for about twenty rotation periods. 
They found no evidence of mode saturation unless the $r$-mode amplitude was
much larger than unity.   \cite{Lind01} carried out numerical simulations of the 
growth of the $r$-modes driven by the current quadrupole radiation reaction
in Newtonian hydrodynamics. To achieve a significant growth 
of the $r$-mode amplitude in a reasonably short computational time, they 
multiplied the radiation reaction force by a factor of 4500,  
decreasing the growth time of the $r$-mode from about 40 s to 10 ms. 
The $r$-mode amplitude $\alpha$ grew to $\approx 3.3$, before shock waves
appeared on the surface of the star and the $r$-mode collapsed. 
They suggested that the nonlinear saturation amplitude of the $r$-modes might be 
set by dissipation of energy in the production of shock waves. 

We note that, with the artificially large radiation reaction, the results 
in \citep{Lind01} assume that no hydrodynamic process takes energy away 
from the $r$-mode in a timescale between 10 ms and 40 s (the artificial growth 
time and the actual physical growth time, respectively). In our previous 
work \citep{paper1}, we showed that $r$-modes of amplitudes unity 
or above were destroyed by a catastrophic decay within this time period.
During the process the $r$-mode motion changed rapidly to a vortical 
motion and strong differential rotation developed.  The mechanism leading to this catastrophic decay is studied in this paper.  

Another approach to study the nonlinear couplings between the $r$-modes 
and other fluid modes is via perturbation theory. In this analytic approach,
one assumes that the system is weakly interactive; and the $r$-mode couples to 
other fluid modes via the lowest order couplings so that higher order effects 
can be ignored. 
Based on a second-order perturbation theory developed by \citet{Schenk}, 
\citet{Arras} proposed a turbulent cascade scenario in which the energy of 
the $r$-mode leaked to a large number of short wavelength inertial modes due to 
three-mode couplings. The coupled modes are in turn damped by viscosity.
They argued that the rotating frame $r$-mode energy could be limited to 
$E_{r-{\rm mode}} \approx 10^{-6} E_{\rm unit}$ for a millisecond rotating star, 
where $E_{\rm unit}= {1\over 2} MR^2\Omega^2$. 
In their normalization, the $r$-mode amplitude ($A_1$ in their notation)
was related to the mode energy by $E_{r-{\rm mode}}=2 E_{\rm unit} A_1^2$. 
Note that the normalization of the $r$-mode amplitude $\alpha$ used by us 
\citep{paper1} is such that $A_1\approx 0.09\alpha$ (see Sec. 8 of \citet{Arras}). 
Hence, in our notation, the maximum $r$-mode amplitude suggested by them is 
$\alpha\approx 8\times 10^{-3}$. 

In a recent second-order perturbative work, \citet{Brink05} have studied the 
evolution of the unstable $r$-mode in a nonlinear network coupled with about 5000 inertial modes. They considered a uniform density, incompressible uniformly rotating 
star. With this simplified model, they calculated the eigenfunctions of the 
inertial modes and the three-mode coupling coefficient $\kappa$ analytically. 
Their study suggested that, while the behavior of the $r$-mode was complicated and depended sensitively on the damping effects of the coupled modes, the $r$-mode amplitude 
would in general be limited to a small amplitude: $c_\alpha \approx 10^{-4}$ in their notation. 
Note that the normalization convention used by \citet{Brink05} is such that the $r$-mode 
energy is given by $E_{r-{\rm mode}}=E_{\rm unit} c_{\alpha}^2$.  
This in turn gives the relation $c_\alpha=\sqrt{2} A_1\approx 0.13 \alpha$, 
and thus the $r$-mode is limited to $\alpha\approx 8\times 10^{-4}$ in our notation.

In the perturbation studies of \citet{Brink05}, 
they found that beyond the critical point $\alpha_{\rm crit}$, energy might 
oscillate between the $r$-mode and the fluid modes it coupled to. After the 
catastrophic decay, the $r$-mode amplitude could re-grow to a value 
comparable to that just before the decay in a short timescale. 
In our numerical simulations with $\alpha$ of order unity, we do not see any such oscillations, 
instead the $r$-mode energy cascades into a large number of fluid modes beyond 
$\alpha_{\rm crit}$. 
We conjecture that such oscillations are not possible in 
fully nonlinear evolutions and realistic neutron stars. This conjecture 
in turns implies that the $r$-mode amplitude is limited to 
$\alpha_{\rm crit}$.

\section{Conclusions and Discussions}

In this paper, we identify a particular three-mode coupling between the 
$r$-mode and the pair of fluid modes responsible for the catastrophic
decay of large-amplitude $r$-modes seen in our previous simulations
\citep{paper1}.  The three modes satisfy a suitable resonance condition to 
high accuracy.  This also demonstrates that hydrodynamic three-mode coupling can 
play an important role in the $r$-mode evolution as first pointed out by \citet{Arras}.
We have also extracted the eigenfunctions of the two daughter modes, and demonstrated 
that they are consistent with having azimuthal mode number $m=-1$. 

While the fully nonlinear simulations are carried out with a large amplitude 
$r$-mode that is obtained by an artificially large gravitational radiation 
reaction, 
we succeed in drawing conclusions on the $r$-mode evolution at smaller 
amplitude through a comparison with a second-order perturbation analysis.  
In particular, we show that the three-mode equations 
(\ref{eq:mode_eq1})-(\ref{eq:mode_eq3}) are valid in the simulation when 
the daughter modes are small, and determine the coupling strength 
$|\kappa_{RAB}|$ of this particular three-mode coupling to be (at least)
$\sim O(10^{-4})$. 
This implies that the two daughter modes will catch up with the $r$-mode at a 
rather small amplitude ($\alpha = \alpha_{\rm crit} \sim O(10^{-2})$), for 
an $r$-mode driven by gravitational radiation 
reaction.  We conclude that at $\alpha \sim O(10^{-2})$ the usual picture of 
the fluid flow of the neutron star dominated by an $r$-mode would no longer 
be valid, with many other modes 
significantly affecting the fluid flow leading to a catastrophic decay.

Our numerical simulations are limited to a spatial 
resolution of $\Delta x \sim 0.3$ km. Fluid modes with wavelengths 
smaller than $\lambda \sim 1$ km cannot be accurately represented. 
It is possible that the couplings of the $r$-mode to a large number of 
shorter wavelength modes can limit $\alpha_{\rm crit}$ to an 
even smaller value (as suggested by \citet{Arras,Brink05}). 
The amplitude for the catastrophic decay $\alpha_{\rm crit}$ 
determined in this paper is as an upper bound.

The study in this paper is carried out with the neutron star described as a 
self-gravitating perfect fluid body with a polytropic equation of state.  
No particular dissipation mechanism is assumed.  This suggests that the 
hydrodynamic behavior we see is of rather general nature.

Finally, we comment on future prospects of numerical investigations of the $r$-mode 
instability. An interesting question that one might ask is whether 
numerical simulations could be used to test the maximal amplitudes inferred in this paper
by starting the $r$-mode evolution directly at $\alpha\sim \alpha_{\rm crit}$ instead of
$\alpha\sim O(1)$ as we have done in our simulations. 
As we have shown in \citet{paper1}, the time it takes for the $r$-mode to reach the 
point of catastrophic decay depends sensitively on the value of $\alpha$. For 
$\alpha \sim \alpha_{\rm crit}$, it would take a time much longer than what 
could be modelled accurately by our finite-differencing code in order to see
a significant growth of the daughter modes and the catastrophic decay of the 
$r$-mode. Nevertheless, it might be possible to use hydrodynamics codes based 
on spectral methods (see, e.g., \citet{Vil02}) to handle such long timescale 
problem, since spectral methods can achieve a much higher numerical accuracy 
in general for smooth fluid flow.


\section*{Acknowledgements}

We thank the anonymous referees for useful comments which improve the 
paper. 
This research is supported in parts by NSF Grant Phy 99-79985 (KDI 
Astrophysics Simulation Collaboratory Project), NSF NRAC MCA93S025, the 
Research Grants Council of the HKSAR, China (Project No. 400803), and the 
NASA AMES Research Center NAS. L.M.L. acknowledges support from a Croucher
Foundation fellowship.


\appendix 

\section{Transformation rules of toroidal modes}
\label{sec:appendix}

In this appendix, we derive the transformation rules for the 
Cartesian velocity components of toroidal modes on the equatorial 
plane. 
An arbitrary fluid displacement vector $\vec{\xi}$ can be 
decomposed in vector spherical harmonics as 
\begin{equation}
\vec{\xi} = \sum_{lm}\left( A_{lm} Y_{lm} \hat{r} + B_{lm}\nabla Y_{lm}
+ C_{lm} \vec{r}\times \nabla Y_{lm} \right) ,
\end{equation}
where $(\hat{r},\hat{\theta},\hat{\phi})$ is the coordinate vector 
basis associated with the spherical polar coordinates $(r,\theta,\phi)$;
$Y_{lm}(\theta,\phi)$ are the standard spherical harmonics;  
the quantities $A_{lm}(r)$, $B_{lm}(r)$, and $C_{lm}(r)$ are functions of 
$r$ only. 
Fluid modes for which $C_{lm}=0$ (for all $l$ and $m$) 
are called spheroidal modes, while those for which $A_{lm}=B_{lm}=0$ are 
called toroidal modes. Here we only consider the latter class of modes. 

The spherical harmonics can be written as 
$Y_{lm} = b_{lm} e^{im\phi} P_l^m(\cos\theta)$, where $b_{lm}$ is a 
constant depending on $l$ and $m$, and $P_l^m$ are the associated 
Legendre polynomials. For a fixed $(l,m)$ component, we have 
the following angular dependence for torodial modes
\begin{equation}
\vec{\xi}_{lm} \sim e^{im\phi}\left[ {im\over \sin\theta}P_l^m(\cos\theta)
\hat{\theta} + \sin\theta {P_l^m}^\prime (\cos\theta) \hat{\phi} \right] ,
\end{equation}
where ${P_l^m}^\prime(x)\equiv dP_l^m/dx$. Considering the physically 
relevant case of real $\vec{\xi}$, the fluid motion of a torodial mode 
on the equatorial plane ($\theta=\pi/2$) is 
\begin{equation}
\vec{\xi}_{lm} \sim m \sin m\phi P_l^m (0) \hat{\theta} 
	- \cos m\phi {P_l^m}^{\prime} (0) \hat{\phi} .
\end{equation}
Using the standard transformation between the spherical basis vectors 
$(\hat{r},\hat{\theta},\hat{\phi})$ and the Cartesian one $(\hat{x},\hat{y},\hat{z})$, 
we obtain the following angular dependence of the Cartesian components 
of $\vec{\xi}_{lm}$ on the equatorial plane:
\begin{equation}
\xi_{lm}^x \sim \cos m\phi \sin\phi {P_l^m}^\prime (0) , 
\end{equation}
\begin{equation}
\xi_{lm}^y \sim - \cos m\phi \cos\phi {P_l^m}^\prime (0) , 
\end{equation}
\begin{equation}
\xi_{lm}^z \sim - m \sin m\phi P_l^m (0) .
\end{equation}
We see that under a $\pi$-rotation ($\phi\rightarrow \phi + \pi$), the Cartesian 
components transform according to 
\begin{equation}
\xi_{lm}^x \rightarrow (-1)^{m+1} \xi_{lm}^x ,
\end{equation}
\begin{equation}
\xi_{lm}^y \rightarrow (-1)^{m+1} \xi_{lm}^y ,
\end{equation}
\begin{equation}
\xi_{lm}^z \rightarrow (-1)^m \xi_{lm}^z . 
\end{equation}


\begin{thebibliography}{}

\bibitem[\protect\citeauthoryear{Andersson}{1998}]{Andersson:1}
Andersson N., 1998, ApJ, 502, 708 


\bibitem[\protect\citeauthoryear{Andersson}{2003}]{Andersson:3}
Andersson N., 2003, Class. Quantum Grav., 20, R105 

 
\bibitem[\protect\citeauthoryear{Andersson \& Kokkotas}{2001}]{Andersson:2}
Andersson N., Kokkotas K. D., 2001, Int. J. Mod. Phys. D, 10, 381  


\bibitem[\protect\citeauthoryear{Arras et al.}{2003}]{Arras}
Arras P., Flanagan E. E., Morsink S. M., Schenk A. K.,
Teukolsky S. A., Wasserman I., 2003, ApJ, 591, 1129 


\bibitem[\protect\citeauthoryear{Brink et al.}{2005}]{Brink05}
Brink J., Teukolsky S. A., Wasserman I., 2005,
Phys. Rev. D, 71, 064029


\bibitem[\protect\citeauthoryear{Casetti et al.}{1997}]{Casetti}
Casetti L., Cerruti-Sola M., Pettini M., Cohen E. G. D., 1997,
Phys. Rev. E, 55, 6566 


\bibitem[\protect\citeauthoryear{Craik}{1985}]{Craik}
Craik A. D., 1985, Wave Interactions and Fluid Flows,
Cambridge University Press 


\bibitem[\protect\citeauthoryear{Friedman \& Lockitch}{2001}]{Friedman:2}
Friedman J. L., Lockitch K. H., 2001, preprint (gr-qc/0102114)


\bibitem[\protect\citeauthoryear{Friedman \& Morsink}{1998}]{Friedman:1}
Friedman J. L., Morsink S. M., 1998, ApJ, 502, 714  
 

\bibitem[\protect\citeauthoryear{Gressman et al.}{2002}]{paper1}
Gressman P., Lin L.-M., Suen W.-M., Stergioulas N., Friedman J. L., 
2002, Phys. Rev. D, 66, 041303(R) 

\bibitem[\protect\citeauthoryear{Lindblom et al.}{2001}]{Lind01}
Lindblom L., Tohline J. E., Vallisneri M., 2001, Phys. Rev. Lett., 
86, 1152

\bibitem[\protect\citeauthoryear{Rezzolla et al.}{2000}]{Rez00}
Rezzolla L., Lamb F. K., Shapiro S. L., 2000, ApJ, 531, L139

\bibitem[\protect\citeauthoryear{S\'{a}}{2004}]{Sa04}
S\'{a} P., 2004, Phys. Rev. D, 69, 084001

\bibitem[\protect\citeauthoryear{Schenk et al.}{2001}]{Schenk}
Schenk A. K., Arras P., Flanagan E. E., Teukolsky S. A., 
Wasserman I., 2001, Phys. Rev. D, 65, 024001 

\bibitem[\protect\citeauthoryear{Stergioulas \& Font}{2001}]{Ste01}
Stergioulas N., Font J. A., 2001, Phys. Rev. Lett., 86, 1148

\bibitem[\protect\citeauthoryear{Villain \& Bonazzola}{2002}]{Vil02}
Villain L., Bonazzola S., 2002, Phys. Rev. D, 66, 123001

\end{thebibliography}
\end{document}